\documentclass[amsmath,
 amssymb,
 aps,prb,
 twocolumn,
 floatfix,
 unsortedaddress,
 superscriptaddress,
 reprint, 
 longbibliography,
 nofootinbib
]{revtex4-2}
\usepackage{graphicx}
\usepackage{dcolumn}
\usepackage{bm}
\usepackage[usenames, dvipsnames, pdftex]{xcolor}
\usepackage[mathscr]{eucal}
\usepackage{ulem} 
\usepackage{tabularx}
 \usepackage[colorlinks, breaklinks, 
 linkcolor=RoyalBlue,
 citecolor=RoyalBlue,
 urlcolor=RoyalBlue]{hyperref}          
 \usepackage[all]{hypcap}
 \usepackage{multirow}
\usepackage{dsfont}
\usepackage{amsmath,amssymb}

\newcommand{\eq}[1]{\begin{align}#1\end{align}}

\newcommand{\msr}{\mathscr}

\newcommand{\delc}{\delta_\mathrm{c}}
\newcommand{\lng}{\langle \ln G \rangle}

\begin{document}
\title{Band-center metal-insulator transition in bond-disordered graphene}
\author{Naba P. Nayak}
\affiliation{Department of Physics, Indian Institute of Technology Bombay, Mumbai 400076, India}
\author{Surajit Sarkar}
\affiliation{Department of Physics, Indian Institute of Technology Bombay, Mumbai 400076, India}
\affiliation{Department of Physics, Concordia University, Montreal, Quebec, Canada}
\author{Kedar Damle}
\affiliation{Department of Theoretical Physics, Tata Institute of Fundamental Research, Mumbai 400005, India}
\author{Soumya Bera}
\affiliation{Department of Physics, Indian Institute of Technology Bombay, Mumbai 400076, India}

\begin{abstract}
We study the transport properties of a tight-binding model of non-interacting fermions with random hopping on the honeycomb lattice. At the particle-hole symmetric chemical potential, the absence of diagonal disorder (random onsite potentials) places the system in the well-studied chiral orthogonal universality class of disordered fermion problems,  which are known to exhibit both a critical metallic phase and a dimerization-induced localized phase. Here, our focus is the behavior of the two-terminal conductance and the Lyapunov spectrum in quasi-1D geometry near the dimerization-driven transition from the metallic to the localized phase. For a staggered dimerization pattern on the square and honeycomb lattices, we find that the renormalized localization length $\xi/M$ ($M$ denotes the width of the sample) and the typical conductance display scaling behavior controlled by a crossover length-scale that diverges with exponent $\nu \approx 1.05(5)$ as the critical point is approached. However, for the plaquette dimerization pattern, we observe a relatively large exponent $\nu \approx 1.55(5)$ revealing an apparent non-universality of the delocalization-localization transition in the BDI symmetry class. 

\end{abstract}
\maketitle
\section{Introduction}

Quenched disorder plays a significant role in determining electronic transport properties particularly when quantum interference enhances its effects and leads to Anderson localization phenomena~\cite{AndersonPRB1958absence,LeeRamaRMP85, EversMirlinRMP}. Such disorder effects are controlled crucially by the symmetries of the disordered Hamiltonian. For instance, in a two-dimensional~(2D) electron gas with potential scattering from random impurities, the sign of the quantum interference correction to the conductivity depends on whether the electronic system has spin-rotation symmetry. As a result of the important role played by such symmetry considerations, our modern understanding of such phenomena relies heavily on a symmetry-based classification of disordered systems~\cite{AltlandPRB1997symmetry, SymmTopoRMP16}.

Tight-binding models of free fermions with real-valued random hopping amplitudes on the nearest-neighbor links of  a bipartite lattice fall in a particularly interesting symmetry class, labeled BDI by Altland and Zirnbauer in their ten-fold classification of disordered systems~\cite{AltlandPRB1997symmetry}. Due to the absence of on-site potential energy terms and the bipartite structure of hopping amplitudes, the free-fermion spectrum in this class is distinguished by the presence of a particle-hole symmetry which guarantees that each eigenstate at energy $+\epsilon$ has a partner at energy $-\epsilon$. The band-center energy $\epsilon = 0$ is thus special.

Within the field-theoretical approach pioneered by Gade and Wegner~\cite{gadeNPB1991replica, gadeNPB1993anderson}, the bare conductivity at the band center receives no quantum corrections in two dimensions, while the density of states develops a characteristic `Gade-Wegner' singularity $\rho(\epsilon) \sim |\epsilon|^{-1} \exp(-\ln^{1/x} (1/|\epsilon|))$ (with $x{=}2$) for energies $|\epsilon|$ smaller than a characteristic crossover energy scale controlled by this conductivity. The conclusion is that such particle-hole symmetric systems can have a critical metallic phase whose low energy properties are characterized by a fixed line within the field-theoretical renormalization group framework.  

As is well-known from the work of Dyson and others~\cite{dysonPR1953dynamics, Th76, Egg78, ZimanPRL82}, the corresponding one-dimensional system has a stronger `Dyson singularity' $\rho(\epsilon) \sim |\epsilon|^{-1} \ln^{-y}(1/|\epsilon|)$ (with $y{=}3$) in the density of states at the band-center. Generalizations to multichannel cases and the nature of the zero energy wavefunctions in both one and 2D cases have also been studied in more recent literature~\cite{BrouwerPRL00, TitovPRB01, DetomsaiDysonPRB16,HastugaiPRB97}.

In the one-dimensional case, this singular behavior can be derived from the properties of an infinite-disorder fixed point of a real-space strong-disorder renormalization group approach~\cite{FisherRG95}. In the 2D scenario, it is difficult to obtain conclusive results from a direct numerical implementation of this strong-disorder renormalization group approach. However, it motivates a closely related strong-disorder analysis~\cite{motrunichPRB2002particle} of the low-energy properties of the critical metallic phase via a connection to optimal defects in a related dimer model. This real-space approach predicts a singularity of the Gade-Wegner form but corrects the associated exponent to $x=3/2$.

This prediction of a modified Gade-Wegner form for the band center singularity in two dimensions has also been confirmed by subsequent work~\cite{MudryPRB2003density, FosterGPh14} that refined the original field theoretical analysis of Gade and Wegner. More recent work has also studied the effects of vacancy disorder in such 2D systems, finding that vacancies lead to a stronger Dyson form of the singularity (albeit with nonuniversal $y$) at intermediate energies before the system crosses over at the lowest energies to the modified Gade-Wegner form~\cite{hafnerPRL2014density, OstrovskyPRL14, sanyalPRL2016vacancy}.

\begin{figure}[!t]
\centering
\includegraphics[width=1\columnwidth]{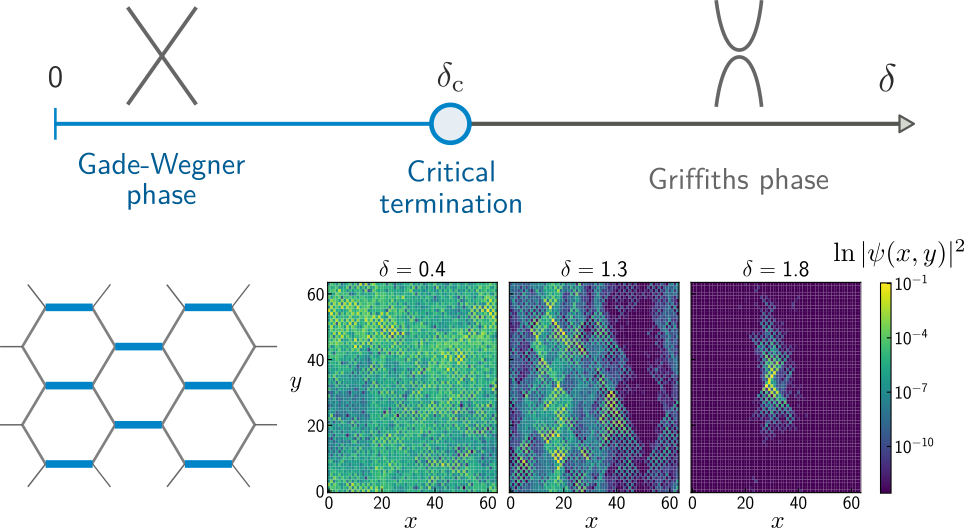}
\caption{Top: Schematic phase diagram as a function of dimerization. Without disorder, there is a dimerization-driven transition from semi-metal to band insulator. With bond disorder, there is a corresponding transition from a critical metal to a Griffiths insulator~\cite{motrunichPRB2002particle}. Bottom left: shows the staggered dimer pattern on a honeycomb flake. Right: $E{=}0$ wavefunction amplitudes $\log |\psi|^2$, for a typical disorder realization of disorder with staggered dimer pattern~\eqref{ham}. The system size is $64\times64$ with periodic boundary conditions in both directions. The wavefunction shows more spreading in the critical metal phase compared to the Griffiths phase as expected. 
}
\label{fig:phasediag}
\end{figure}

The real space approach of \textcite{motrunichPRB2002particle} also predicts that such two-dimensional systems can realize, in addition to the critical metallic phase, a localized Griffiths phase with a weaker non-universal power-law divergence $\rho(\epsilon) \sim |\epsilon|^{-1+2/z}$ (with nonuniversal $z$) of the density of states near the band center. As noted in \textcite{motrunichPRB2002particle}, such a localized phase can be established by strong enough dimerization in the values of the hopping amplitudes (see Fig.~\ref{fig:phasediag}). While both the critical metallic phase and the localized Griffiths phase have been discussed extensively in the literature, the transition between the two has received much less attention~\cite{motrunichPRB2002particle, BocquetPRB03, KonigPRB12}.

Here, we focus on the transport properties of the 2D sample in the vicinity of this transition via numerical studies of both the two-terminal Landauer conductance and the full Lyapunov spectrum $\lambda$. 
The transfer matrix and the Lyapunov spectrum studies provide complementary evidence in favour of or against particular scaling results that are obtained by the renormalization group approach.
For small dimerization, the density $\rho(\lambda)$ of Lyapunov exponents $\lambda$ in quasi-1D geometry is nonzero at $\lambda=0$. This corresponds~\cite{ChalkerAPS1993scattering} to a nonzero conductivity. 
In this phase, we find that the conductivity extracted from the typical two-terminal conductance is roughly independent of the sample geometry and consistent with the conductivity obtained from the Lyapunov density $\rho(0)$, indicating a metallic phase. 
With increasing dimerization, the Lyapunov density of states develops a gap around $\lambda=0$, signaling the transition to a gapped insulating phase at a critical dimerization strength $\delta_c$. A qualitative phase diagram is shown in Fig.~\ref{fig:phasediag}. 

In the gapped insulating phase close to $\delta_c$, we perform a finite size scaling analysis of the renormalized localization length $\xi/M$ (where $M \gg 1$ is the width of the quasi-1D sample whose length $L \gg M$). For staggered dimerization on both the square and honeycomb lattice, we argue that the rigid shift~\cite{motrunichPRB2002particle} of the Lyapunov spectrum, characteristic of staggered dimerization, implies that $\xi/M$ diverges as $\sim (\delta - \delta_c)^{-\nu}$ with $\nu = 1$ as $\delta$ approaches $\delta_c$ from above. The value of $\nu$ we obtain from both the Lyapunov spectrum study and direct measurement of the typical two-terminal conductance in the wide-sample geometry is consistent with this prediction within the numerical errors associated with these calculations. A calculation of the transport properties in the direction perpendicular to the strong bonds of the staggered dimerization pattern also yields the same value of $\nu$ within numerical uncertainties. However, on the honeycomb lattice, for a more symmetric plaquette dimerization that does not have this rigidity property, we observe a larger exponent $\nu \approx 1.55(5)$. This apparent non-universality in the value of the exponent is one of our main findings.

Intriguingly,  the larger value of $\nu$ that we find for plaquette dimerization on the honeycomb lattice is consistent, within our numerical errors, with a recent prediction~\cite{Kracher22} of the same exponent in the closely related problem of random bipartite hopping with {\em complex} hopping amplitudes on the square lattice with the same kind of plaquette dimerization. This falls in the Altland-Zirnbauer class  AIII, while the systems we study are in class BDI. Since this study also used a plaquette dimerization pattern for their numerical work, it would be interesting in follow-up work to ask if a similar non-universality is also present in the problem with complex hopping amplitudes and explore this apparent nonuniversality within the framework of the theory developed in Ref.~\onlinecite{Kracher22}.

\section{Model and observables} \label{model}

We consider a model of free fermions hopping on the honeycomb lattice, and an analogous model on the square lattice. In this model, the real-valued nearest-neighbor hopping amplitudes are independent random variables sampled from a uniform distribution. For conductance calculation, we take a clean one-dimensional lead attached to the disordered sample.  The Hamiltonian of the system thus reads
\begin{eqnarray}
\hat{\msr{H}} = - \sum_{\langle i j \rangle } t_{ij}(c_i^{\dagger}c_j + \mathrm{h.c.}),
\label{ham}
\end{eqnarray}
where $c_i$($c_i^{\dagger}$) is the fermion annihilation (creation) operator at site $i$, and the sum is over all nearest-neighbor links $\langle i j \rangle$.
We define a dimerization pattern by choosing the hopping strength of selected ``dimerized'' bonds to be independent random variables uniformly distributed in the range $[0,e^\delta]$, while the hopping amplitudes on other ``non-dimerized'' bonds are random variables drawn uniformly in the range $[0,1]$. The parameter $\delta$ thus represents a dimerization in the mean value of the hopping amplitudes. Since the ratio of the width of the distribution to its mean remains the same, tuning $\delta$ changes the dimerization without changing the strength of the bare disorder. {In all our numerical simulations, typically, we average over ${\sim}10^3$ disorder configurations for the correlation length calculations, and for conductance typically  we average over ${\sim}10^4$ samples unless specified otherwise.}

We study both the staggered and plaquette dimerization patterns (described below) on the honeycomb lattice, and the staggered pattern on the square lattice to access the dimerization-driven transitions from metallic to localized behavior in this model on both lattices at $E=0$. The staggered anisotropic pattern was chosen as a representative of patterns that break the rotational symmetry between the transverse direction and the direction of the current and leads to a rigid shift of the Lyapunov spectrum. The other plaquette pattern was chosen because it maintains the rotational symmetry and does not lead to a rigid shift of the Lyapunov spectrum. Further, note 
that the staggered dimerization pattern can choose any one out of the three equivalent bond orientations and enhance the corresponding bond strengths. Given the symmetry of the honeycomb lattice, all these choices are equivalent, and the only physical distinction then is in the direction of the current flow relative to the chosen orientation for the staggered pattern.

\begin{figure*}[thb]
\centering
\includegraphics[width=1\textwidth]{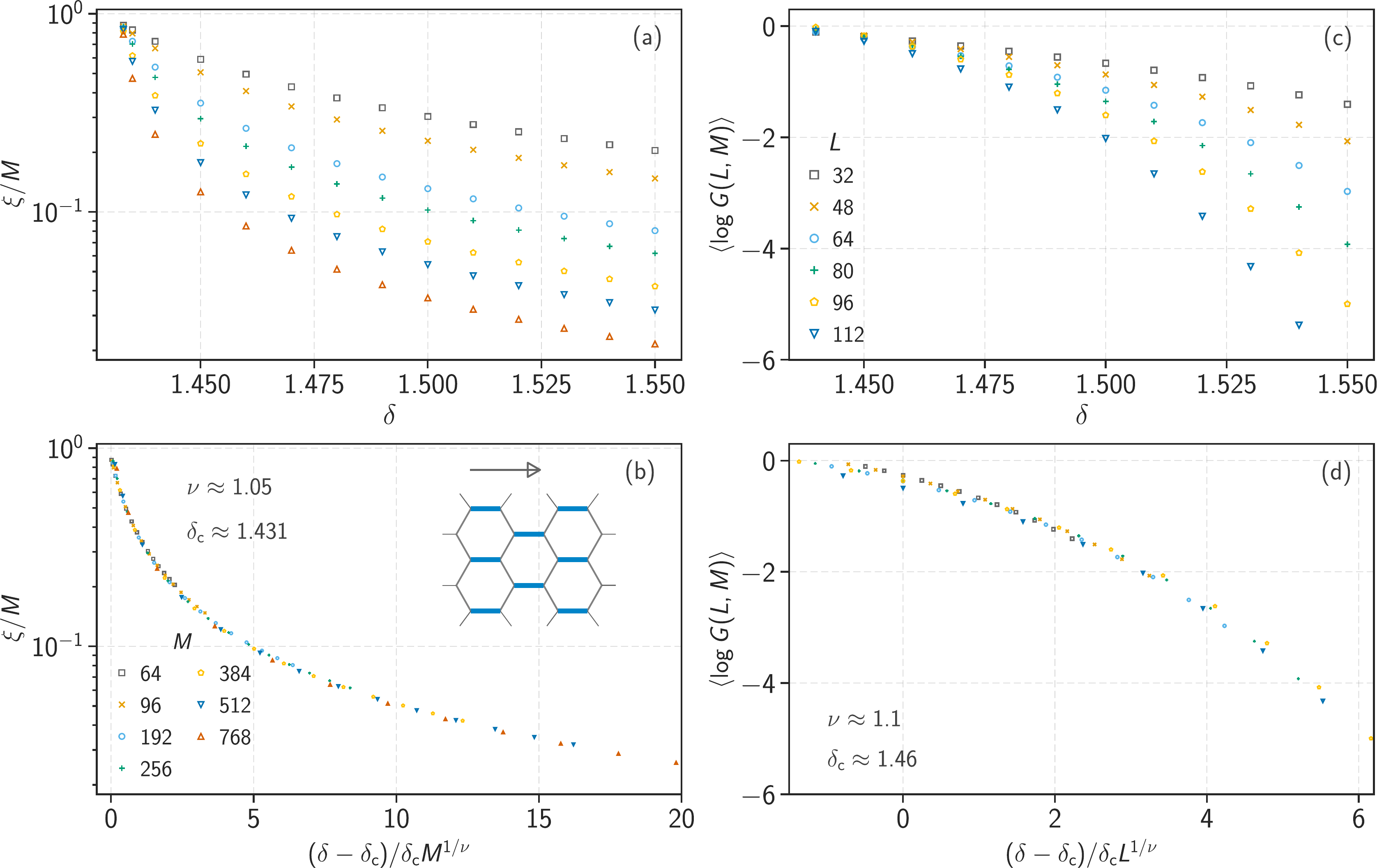}
\caption{ Results for staggered dimerization pattern on arm-chair graphene shown in Fig~\ref{fig:loclaizedphase_stagger_gphx}(b) inset. The normalized localization length $\xi/M$ for a transfer along the direction of the arrow of the lattice as a function of dimerization $\delta$ for different values of width $M=\{64-768\}$ and length $L{=}10^5$ in the localized phase. (b) Shows the corresponding finite size data collapse with critical exponent $\nu\approx1.05$, and a critical dimerization strength $\delc \approx 1.431$.   (c) The $\lng$ is calculated within the two terminal setup for the same lattice with 1D clean lattice leads connected along the direction of the arrow in the localized phase for various system lengths $L=\{32-112\}$ for $L/M\sim1/4$. (d)~The corresponding approximate scaling collapse is shown for $\lng$ which yields $\nu\approx1.1$.}
\label{fig:loclaizedphase_stagger_gphx}
\end{figure*}

In the staggered dimerization pattern on the honeycomb lattice (shown in Fig.~\ref{fig:loclaizedphase_stagger_gphx}(b) inset) the horizontal bonds along the direction of the arrow are dimerized, while along the other direction the bonds are non-dimerized in the way defined above.
The plaquettes dimerization pattern on the honeycomb lattice (shown in fig~\ref{fig:loclaizedphase_plaquettes}(b) inset) is a more symmetric pattern that does not single out a particular direction as in the staggered pattern. In this pattern, the strong ``dimerized'' bonds are ordered at the three-sublattice wavevector of the underlying triangular Bravais lattice. On the square lattice, we mainly study the staggered pattern, in which the strong ``dimerized'' bonds are all of one orientation and arranged in a pattern corresponding to wavevector $(\pi, \pi)$. The plaquette dimerization pattern on the square lattice on the other hand has strong ``dimerized'' bonds that form perimeters of elementary plaquettes that are ordered at wavevectors $(\pi, 0)$ and $(0,\pi)$, and has been studied in Ref.~\cite{Kracher22} with complex hopping amplitudes.  

\begin{figure}[!tb]
\centering
\includegraphics[width=1\columnwidth]{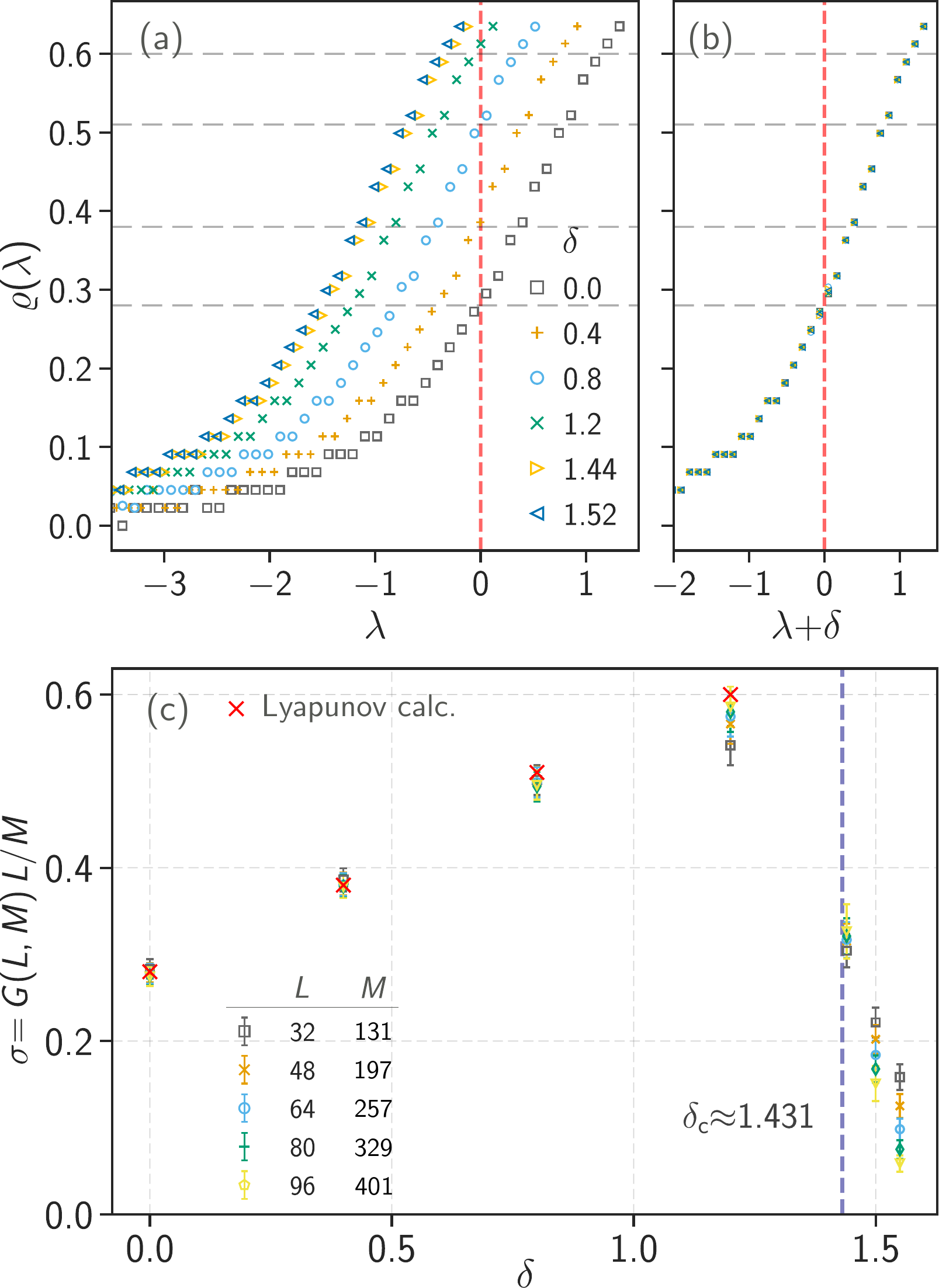}
\caption{(a)~The Lyapunov spectral density $\rho(\lambda)$ for A-sublattice transfer along the direction of the arrow in the staggered graphene lattice as shown in Fig.~\ref{fig:loclaizedphase_stagger_gphx}(b) inset. With increasing $\delta$, the spectral density shifts rigidly, and the horizontal dotted lines indicate the value where the density curves cut the $\lambda=0$ axis, i.e. $\rho(\lambda=0)$ and this gives an estimate of the conductivity (upto a trivial constant) of the sample in the metal phase~\cite{ChalkerAPS1993scattering}.
The right panel shows the collapse of all curves with the rigid shift by $\delta$.  (c)~Shows the conductivity $\sigma=G(L,M)L/M$ as a function of $\delta$ estimated from the Lyapunov density of states in quasi-$1$D geometry and the scattering matrix formalism for a wide sample $L/M\sim1/4$~(see further details in the text). One-dimensional lattice leads are used for the conductance calculation. }
\label{fig:lyp_dos_A_sub}
\end{figure}
%
%
\subsection{Lyapunov analysis} 
The standard Lyapunov spectrum $\lambda_i$ of the transfer matrix $T=\prod_n T_n$ ($T_n$ is the transfer matrix for the $n$-th slice) for a quasi-$1$D geometry of width $M$ is used to study the metal-insulator transition (MIT)~\cite{markos2006numerical}.
In the quasi-1D limit, the inverse of the smallest Lyapunov exponent in the limit of large $L$ in the localized phase defines the localization length $\xi=1/\lambda_\text{min}$~\cite{ChalkerAPS1993scattering}. The normalized localization length $\xi/M$~\cite{ChalkerAPS1993scattering} is expected to exhibit finite size scaling across an MIT, allowing one to extract the exponent $\nu$ from a finite-size scaling analysis of $\xi$ approaching the transition from the localized side. 
Moreover, the limiting spectral density of Lyapunov exponents \{$\lambda_i$\} in such a quasi-1D geometry, 
\begin{equation}
   \rho{(\lambda)} = \lim_{M \rightarrow \infty} 1/M \sum_i \delta(\lambda-\lambda_i),
   \label{lyp_dos}
\end{equation}
contains information about the transport properties of the sample (here the width $M$ of the sample is assumed to increase while maintaining the quasi-1D geometry with $L \gg M$).
In particular, the $\lambda=0$ density of states $\rho(\lambda{=}0)$ is finite in a metal and proportional to the conductivity, while it is strictly zero in an insulating phase~\cite{ChalkerAPS1993scattering}.  
Therefore, at a transition to a localized insulating phase, we expect the Lyapunov spectral density to develop a gap in the spectrum at $\lambda=0$. 
In our calculation, we use the fact that the Hamiltonian has chiral symmetry. 
Due to this, the wavefunctions at $E=0$ can be chosen to have support only on one sublattice, and the transfer of such a wavefunction from one side to the other can be performed just on this sublattice; this decoupling allows a trivial factor of two increases in the system width $M$~\cite{motrunichPRB2002particle,montrunicsThesis}, since one can assemble the full Lyapunov spectrum from such a calculation for just one sublattice. The full spectrum $\rho(\lambda)$ is symmetric about $\lambda = 0$, and this symmetry implies that a Lyapunov mode at $+\lambda$ on one sublattice has a partner at $-\lambda$ on the other sublattice, allowing a reconstruction of the full spectrum from this sublattice calculation.

As has been emphasized earlier~\cite{motrunichPRB2002particle,montrunicsThesis}, the staggered dimerization pattern leads to an interesting simplification: The two sublattice Lyapunov spectra at $E=0$ shift rigidly in opposite directions with increasing $\delta$, all the while maintaining the $\lambda \rightarrow -\lambda$ symmetry of the full spectrum~(see Appendix~\ref{app1} for more details). It is therefore possible to determine the critical value $\delta_c$ at which the metal gives way to the insulator simply by knowing the spectrum for one sublattice at $\delta = 0$. For $\delta > \delta_c$ in the insulator, this ridigity also implies that $\lambda_{\rm min} \sim (\delta - \delta_c)$, implying that $\xi/M  \sim 1/(M(\delta -\delta_c)$. This argument suggests that the correlation length exponent $\nu$ takes on the value $\nu =1$ for an MIT driven by such a rigid shift in the sublattice Lyapunov spectrum. It is interesting to note that such an argument is also valid for complex hoppings with staggered dimerization pattern, but we have not studied it further here. 

Below we explore the validity of this argument and check if this also controls the scaling of the two-terminal conductance calculated from {the scattering wavefunctions} (described below). We also ask if a generic transition, not driven by dimerization patterns that involve a rigid shift, has a different value of $\nu$ characterizing its critical behavior.

\subsection{Two terminal conductance}

The two terminal linear conductance at zero temperature is calculated using the Landauer formula
$$ G= \frac{e^2}{h} \sum_{nm}  |S_{nm}|^2, $$
where $G$ is the dimensionless conductance (in units of $e^2/h$) and $S_{nm}$ is the scattering matrix element between the scattering channel $n$ and $m$ obtained in terms of the zero energy wavefunctions.
It relates electrical conductance to the total transmission probability of electron waves through a region of random scatters. As is well known, a closely related analysis~\cite{BarangerStonePRB89} in terms of Green functions also leads to the  Landauer formula, which relates electrical conductance to the total transmission probability of electron waves through a region of random scatters; in this sense, our results are expected to be equivalent to those obtained from explicit use of the two-terminal Landauer formula.

We used open source {\tt Kwant} package~\cite{Groth2014Kwant}, a wavefunction-based method for the conductivity calculation. The `effective' disorder in the system is large, and thus the conductivity is small $\mathcal{O}(1)$. To avoid small numbers we use a wide geometry $M >  L$, which has more conducting channels and ensures a larger value of the conductance. This is useful because the divergent density of states at zero energy gives rise to numerical instability in the opposite limit of a quasi-1D sample. 
In all our conductance calculations we use one-dimensional lattice leads which allow having a finite density of state in the lead at $E=0$. 
The sample width $M$ is chosen in such a way that the clean band structure with armchair edges has linear energy dispersion.

\begin{figure}[!tbp]
    \centering
    \includegraphics[width=1\columnwidth]{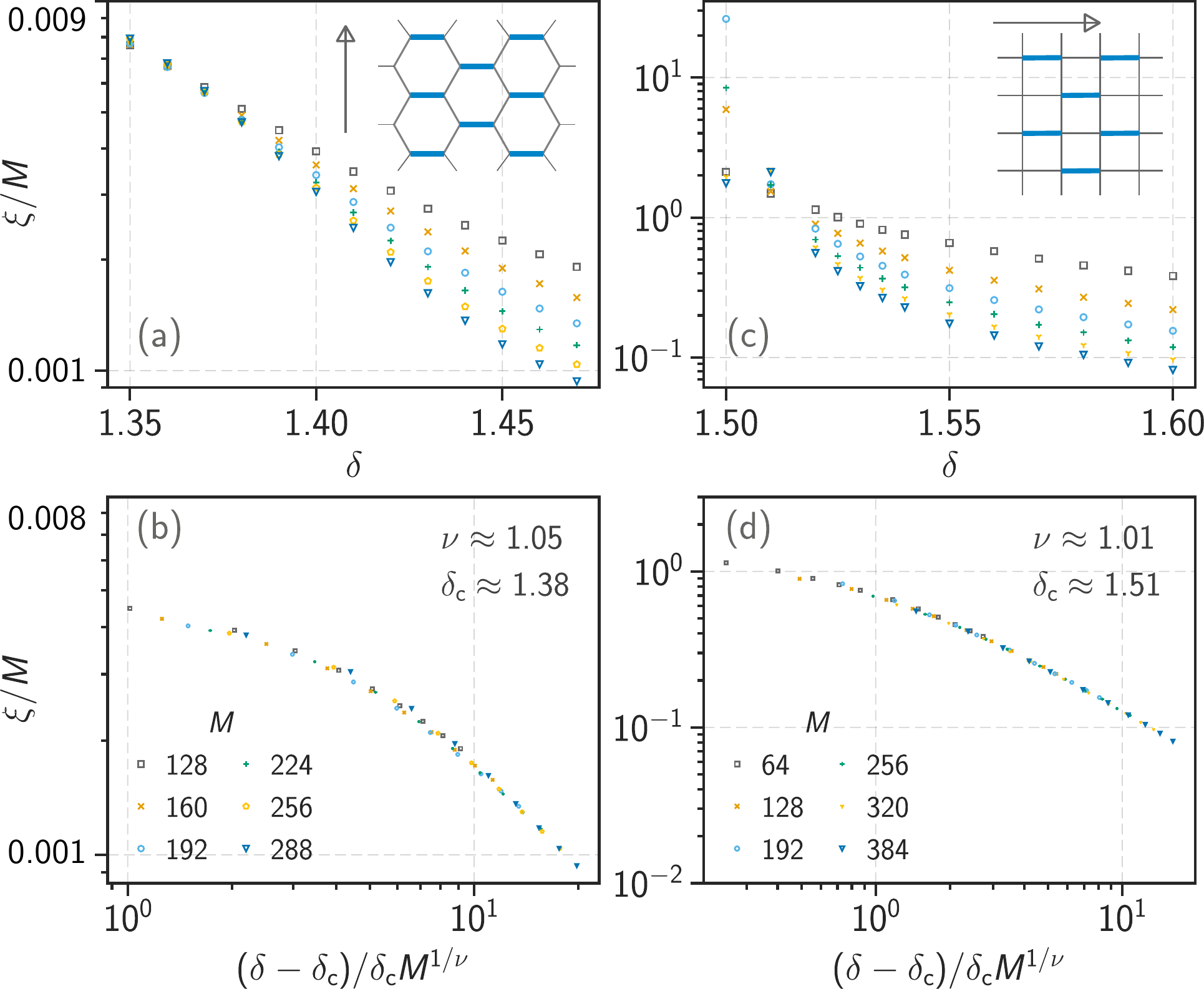}
   \caption{(a) Renormalized localization length as a function of the dimerization strength $\delta$ for staggered dimerization pattern on the honeycomb lattice in the direction of the arrow as shown in the inset. (b) Shows the scaling collapse, which yields $\nu\approx 1.05(5)$.  (c)~The renormalized localization length $\xi/M$ for stagger dimerization pattern on a square lattice for a transfer along the direction of the arrow. Details are similar to (a). (d) The corresponding data collapse yields a  critical exponent $\nu\approx 1.01(3)$. }
   \label{localizedphase_stagger_gphy}
\end{figure}
In a generic Anderson delocalization-localization transition, the typical conductance $\exp(\lng )$ acts as an order parameter and displays scaling behavior~\cite{SlevinPRL2001reconcile}. 
With this in mind, we study the $L$ and $\delta$ dependence of $\lng$, and also we monitor the probability distribution of $\ln G$ for a range of $L$ and $\delta$ in the metallic phase. We obtain data for the staggered dimerization patterns on both the square and honeycomb lattice, and honeycomb lattice data for the plaquette dimerization pattern (Fig.~\ref{fig:loclaizedphase_stagger_gphx}, Fig.~\ref{localizedphase_stagger_gphy}, and insets).
\section{Numerical Results \label{numerics}}

\subsection{Staggered dimerization on the honeycomb lattice}

\begin{figure*}[thb]
\centering
\includegraphics[width=1\textwidth]{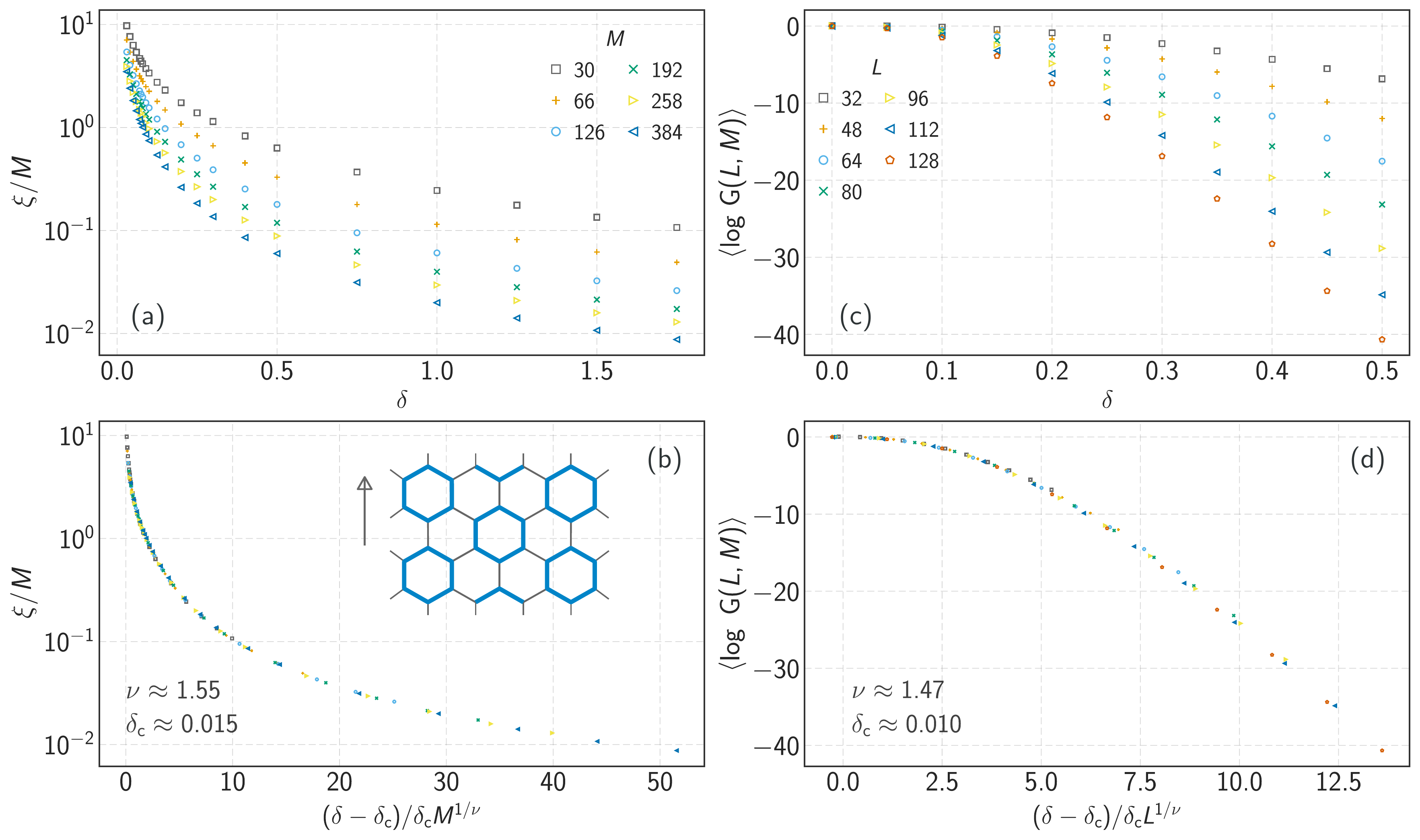}
\caption{(a) Results for plaquette dimerization pattern on arm-chair graphene as shown in Fig~\ref{fig:loclaizedphase_plaquettes}(b) inset. The normalized localization length $\xi/M$ as a function of dimerization $\delta$ for different values of width $M=\{30-384\}$ and length $L{=}10^5$ in the localized phase. (b) Shows the corresponding finite size data collapse with critical exponent $\nu\approx1.55$, and a critical dimerization strength $\delc \approx 0.015$. (c) The $\lng$  is calculated within the two terminal setups for the same lattice with 1D leads connected along the direction of the arrow in the localized phase for various system lengths $L=\{32-128\}$ for $L/M\sim1/4$. (d)~The corresponding approximate scaling collapse is shown for $\lng$ which yields $\nu\approx1.47$.}
\label{fig:loclaizedphase_plaquettes}
\end{figure*}


\paragraph*{Lyapunov spectrum and $\delc$:}
The Lyapunov spectral density~\eqref{lyp_dos} for $A$ sublattice transfer along the direction of the arrow in the staggered quasi-1D sample (shown in fig~\ref{fig:loclaizedphase_stagger_gphx}(b) inset) of width $M=768$ is shown in Fig.~\ref{fig:lyp_dos_A_sub}(a). The curves show a finite density at $\lambda=0$ in the metallic phase. 

Increasing $\delta$ further creates a finite gap around the center i.e. zero spectral density at $\lambda = 0$. This marks the onset of the localized phase of the system and that allows a precise determination of $\delc$. The estimated critical disorder strength is found to be $\delta_c\approx1.431$. In this case, the spectrum does shift rigidly, consistent with the general argument made earlier. 

In the Griffiths insulator phase, we see that $\xi/M$ decreases with increasing $\delta-\delta_c$ and $M$ showing increasing localization as seen in Fig.~\ref{fig:loclaizedphase_stagger_gphx}(a). 
A one-parameter scaling of $\xi/M$ is performed with following scaling form $\xi/M{=}\mathcal{F}\left(\chi(w)M^{1/\nu},\phi(w) M^{y} \right)$ (see Appendix~\ref{scaling_appendix} for more details), where $y$ is the irrelevant scaling exponent.  The corresponding data collapse is shown in Fig.~\ref{fig:loclaizedphase_stagger_gphx}(b). The estimated critical disorder strength from the analysis is $\delc{=}1.431(2)$, which is consistent with the critical dimerization value extracted from the Lyapunov density of state (see in Fig.~\ref{fig:lyp_dos_A_sub}). The localization length exponent is found to be $\nu{=}1.05(5)$, consistent with the theoretical prediction of $\nu = 1$ for cases when the dimerization leads to a rigid shift in the sublattice Lyapunov spectrum.

\paragraph*{Conductance calculation in localized phase}
We compute $\lng$ via {scaterring matrix formulation} as described earlier. In the localized phase, this provides an independent window to these localization properties.

We find that the conductance is exponentially small for $\delta > \delta_c$, {\em i.e.} $G\ll 1$ and decreasing with system length $L\gg\xi$ as well as with increasing dimerization $\delta$ 
 as shown in Fig.~\ref{fig:loclaizedphase_stagger_gphx}(c).
We investigate the dependence on $\delta$ and $M$ via a finite size scaling analysis at fixed $L/M$ for wide samples as shown in Fig.~\ref{fig:loclaizedphase_stagger_gphx}(d). 
The scaling collapse gives a critical disorder to be $\delta_c {=} 1.46(2)$ and the exponent $\nu = 1.10(4)$. The marginally higher critical parameters can be attributed to a slightly different geometry $L/M\approx1/4$ used for the numerical simulation. Within the available computational resources, we could not directly probe the true 2D limit $L/M {=} 1$ as it would require simulation of large sample sizes.  The estimate of $\nu$ obtained in this way is also very close to the value obtained from the Lyapunov exponent analysis; indeed the theoretical prediction of $\nu =1$ is just outside the error bars of this numerical estimate.

\paragraph*{Additional confirmation of scaling behavior for staggered dimerization:}
To explore this further, we also determine $\nu$ using a Lyapunov spectrum analysis for transfer in the perpendicular direction i.e., along the zigzag edge of the honeycomb lattice (shown in Fig.~\ref{localizedphase_stagger_gphy}(a) inset). The results are shown in Fig.~\ref{localizedphase_stagger_gphy}(a). A one-parameter scaling collapse gives a localization length exponent of $\nu\approx 1.05$ in agreement with the transfer along the other direction. The critical disorder strength is estimated to be $\delta_c\approx 1.38$. 
The  smaller critical disorder strength in this direction could be related to the strong anisotropy of the bond arrangement. We also observed (data not shown) an extremely slow flow towards the larger $\delta_\text{c}\sim 1.43$ with increasing system sizes $M$.

Additionally, we also verify our results for staggered dimer patterns on a square lattice. The main difference here is that the density of states at $E=0$ is finite in the clean limit compared to the Dirac density of states.
The results for the transfer matrix are shown in Fig.~\ref{localizedphase_stagger_gphy}. A finite size analysis again predicts the exponent to be $\nu\approx 1.01$ in agreement with all our previous estimates of staggered dimer patterns on the honeycomb lattice. Again, we attribute this value of $\nu$ to the fact that the Lyapunov spectrum on each sublattice shifts rigidly with $\delta$ for this kind of staggered dimerization.

\begin{figure*}[btp]
    \centering
    \includegraphics[width=1\textwidth]{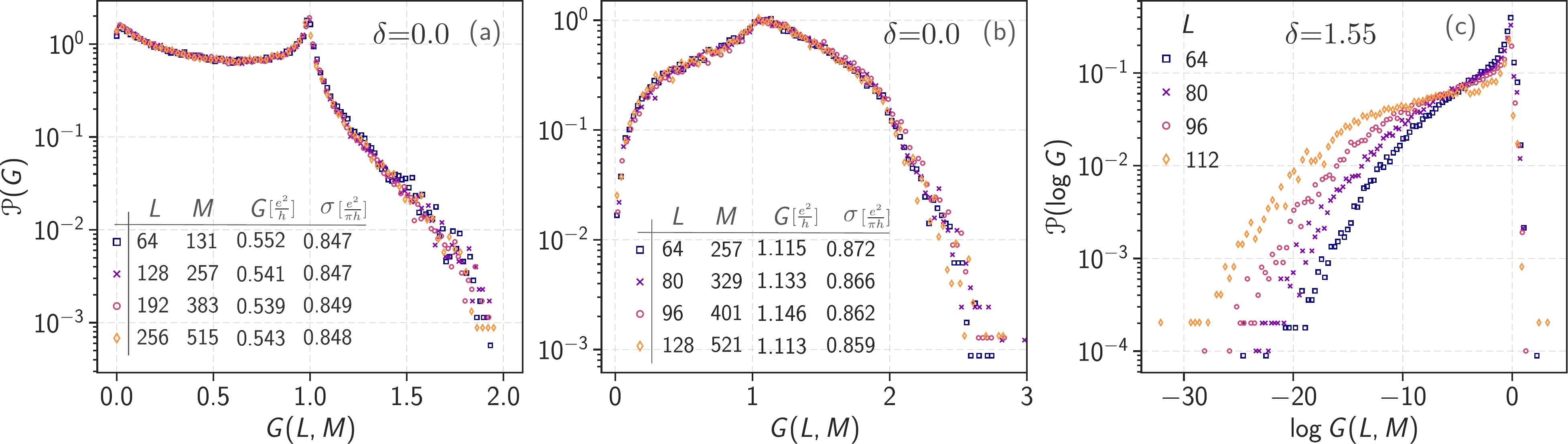}
   \caption{(a)-(b) Conductance distribution $\msr{P}(G)$ at $\delta{=}0$ is calculated within the two terminal geometry for the finite staggered honeycomb sample shown in Fig.~\ref{fig:loclaizedphase_stagger_gphx}(b) inset, with two different aspect ratios $L/M \sim 1/2, 1/4$. The corresponding $L, M$, and the average conductance $G[e^2/h]$ along with the mean conductivity $\sigma[e^2/(\pi h)]\approx 0.85(2)$ are indicated in the legend. (c) Shows the $\ln G$ distribution in the localized phase $\delta{=}1.55$ for different sample length $L=\{64-112\}$. Open armchair boundary is used for these calculations. } 
    \label{fig:g_dist_staggered}
\end{figure*}

\paragraph*{Critical metal phase:}
Following \textcite{ChalkerAPS1993scattering} we estimate the average conductivity in the critical metal phase $\sigma = G(L,M) L/M = b \rho(0)$, where $b$ is an order one number, and $\rho(0)$ is the Lyapunov density, which is shown in Fig.~\ref{fig:lyp_dos_A_sub}.
The two terminal conductance calculated in the wide sample limit $L/M \ll 1$ is also shown in Fig.~\ref{fig:lyp_dos_A_sub} as a function of dimerization $\delta$ and compares well to the estimate obtained above from the Lyapunov spectrum. 
In the metallic phase, it increases with $\delta$ until the metal-insulator transition point is reached, perhaps reflecting the fact that the strengthened dimerization facilitates transport in the measured direction.

\subsection{Plaquette dimer configuration}
%
The plaquette dimerization pattern on the honeycomb lattice is shown in Fig.~\ref{fig:loclaizedphase_plaquettes}(b) inset. This pattern does not single out a particular direction of the lattice as in the case of a staggered pattern.
The plaquette dimerization also drives a dimerization-driven critical metal-insulator transition, however, with a larger localization length exponent, which could point towards a different fixed point.  

The normalized localization length $\xi/M$ and log-conductance $\lng$ data is shown in Fig~\ref{fig:loclaizedphase_plaquettes}(a) and in Fig~\ref{fig:loclaizedphase_plaquettes}(c) respectively. The finite-size analysis of both of our observables (Fig~\ref{fig:loclaizedphase_plaquettes}(b) and Fig~\ref{fig:loclaizedphase_plaquettes}(d)) points towards a higher localization length exponent in this dimer configuration. The localization length predicts the exponent to be $\nu\approx1.55(2)$ which agrees within error bars to the conductance data which predicts $\nu\approx 1.4(1)$. As noted earlier, recent work in the other chiral class AIII indeed predicts a larger exponent $\nu\approx1.55(1)$ with a similar dimer pattern on the square lattice with complex hopping~\cite{Kracher22}.

\subsection{Conductance distribution} 
In Fig.~\ref{fig:g_dist_staggered} we show the distribution of the conductance $G(L,M)$ for the honeycomb lattice, in the metallic phase at $\delta=0$.
The data is shown for two different aspect ratios $L/M=1/2, 1/4$ Fig.~\ref{fig:g_dist_staggered}(a,b). 
For both the aspect ratios the width of the $\mathscr{P}(G)$ is independent of the sample $L$, indicating its scale-invariant properties. 
%
%
While the average conductance of course depends on the sample geometry, the conductivity $\sigma{\approx}0.85$  (in the units of $e^2/(\pi h)$) for $\delta{=}0$ is independent of the sample geometry indicating that the value is close to its true 2D limit.
The distribution $\mathscr{P}(G)$ has a long tail with a singularity at $G\approx 1$ (in the units of $e^2/h$), which is a reminiscence of the critical conductance distribution at a 3D metal-insulator transition~\cite{MarkosPRB02}.

In the localized insulating phase, the mean conductance is small $G(L,M)\ll1$; therefore we show the log-conductance $\ln G$ distribution in Fig.~\ref{fig:g_dist_staggered}(c) at $\delta=1.55$, which is quite close to $\delta_c$. Deep in the localized phase, the conductance becomes extremely small and suffers from numerical instability. The $\msr{P}(\ln G)$ is far from a normal Gaussian form, which one would expect in the localized phase~\cite{markos2006numerical}. However, we observe the peak at $G\approx 1$ decreases rapidly with $L$ in this regime and the mean conductance becomes smaller. The scaling analysis of the mean $G$ has already been presented in Fig.~\ref{fig:loclaizedphase_stagger_gphx}.

\section{Discussion and Outlook\label{summary}}
We present a numerical study of transport properties in a bond-disordered tight-binding model of honeycomb and square lattice. The disordered version of the phase transition is allied with the opening of a band gap in the clean model due to dimerization.

From our results for the critical behavior of the mean log-conductance and normalized localization length, we find an apparent non-universality of critical exponents in the BDI symmetry class. In our analysis we estimated it to be $\nu \approx 1.05(5)$ for staggered dimer pattern driven localization transition, and the critical exponent to be $\nu\approx 1.55(5)$ for plaquette dimer configuration. Important to note that with particle-hole symmetry, the localization transition is actually preceded by the critical metal phase that we have discussed here, and not by the weak localized phase that is usually observed in 2D standard Anderson transition. The critical phase itself is a strongly disordered phase as signalled by small conductivity $\sigma[e^2/\pi h] \lesssim 1$.


At the critical point, we find that the distribution of the conductance is scale invariant. In the range of sizes accessible to our study, this scale-invariance remains approximately valid for a range of $\delta$ on the metallic side of the transition. 

In the immediate vicinity of the transition in the insulating side, the distribution shows a significant deviation from the critical distribution. However, the expected log-gaussian distribution of log-conductance is yet to develop fully in our calculation.

Recently, similar transport statistics have been studied in quasi-1$d$ armchair graphene with bond disorder~\cite{Burnell21}. 
In this work, the focus was to understand the crossover phenomenon of the quasi-localized~(chiral) critical point to an exponentially localized regime by two parameters scaling with respect to the energy and the system sizes. 
On the contrary, we are in the 2D limit, where the shape of the conductance fluctuations remains log-Gaussian across the phase transition unlike in the 1D model. 
In particular, we observe system size-independent conductance fluctuations close to the critical point, which is absent in the quasi-1D limit. 

In the future, it would be relevant to investigate the effects of temperature and interaction on the chiral symmetric critical point. Exploring the stability of the phase transition and the interaction effects on the universality of the exponent would be important to investigate.
Similarly, it would be important to understand the origin of such non-universality in the exponent for different dimerization patterns, possibly, using the field theoretical framework that has been developed in Ref.~\cite{Kracher22} and in particular, when the symmetry of lattice rotations is broken by the dimerization pattern.

\section{ACKNOWLEDGMENTS}
We are grateful to Sasha Mirlin for pointing out a crucial numerical error in an earlier version of the manuscript, and for several discussions. SB acknowledges support from SERB-DST, India, through Matrics (No. MTR/2019/000566), and MPG for funding through the Max Planck Partner Group at IITB. NN would like to thank DST-INSPIRE fellowship No. IF- 190078 for funding. KD was supported at the TIFR by DAE, India, and in part by a J.C. Bose Fellowship~(JCB/2020/000047) of SERB, DST India, and by the Infosys-Chandrasekharan Random Geometry Center~(TIFR).
\bibliography{ref}
\appendix
\section{Transfer Matrix and Rigid Spectrum}
\label{app1}
\begin{figure}
    \includegraphics[width=0.4\textwidth]{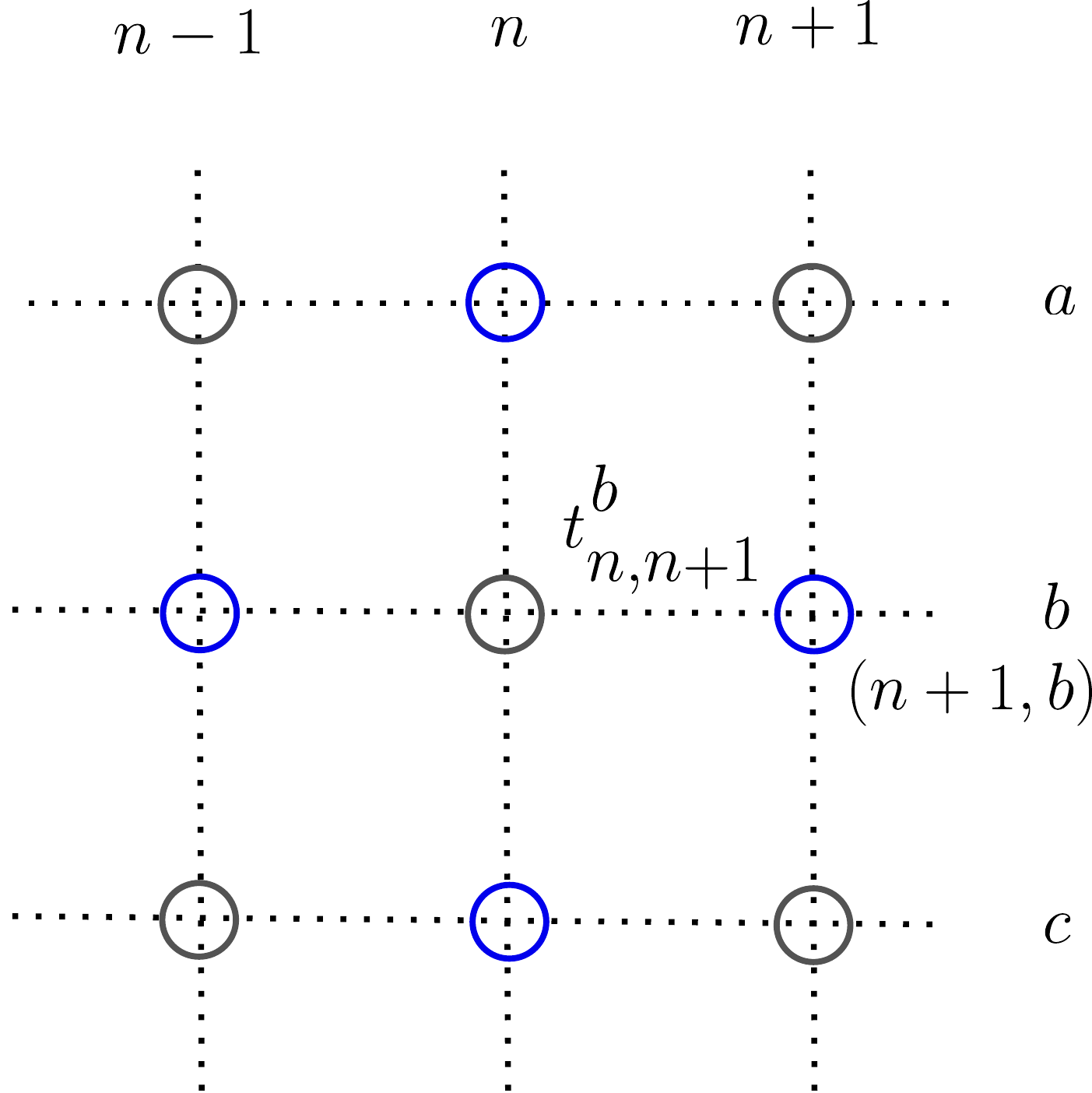}
    \label{rigid_shift}
\caption{Shows the  2D square lattice geometry for transfer matrix calculation. Blue and gray circles represent two different sub-lattices in this bipartite lattice.  
}
\end{figure}
Here we discuss the rigid shift in Lyapunov spectrum observed in Fig.~\ref{fig:lyp_dos_A_sub} in detail. Figure~\ref{rigid_shift} shows a part of a 2D square lattice with vertical slices denoted by $(n-1,n,n+1)$ and horizontal slices are denoted by $(a,b,c)$. The hopping $t^b_{n,n+1}$ defines a connection between the lattice points $(n,b)$ and $(n+1,b)$.



The wave-amplitudes at the slice $n+1$ at $E=0$ are given by the following transfer matrix~\cite{markos2006numerical}
\eq{
\left(
\begin{array}{c}
\psi_{n+1} \\
\psi_{n}
\end{array}
\right)
=\left( \begin{array}{cc}
\frac{-h_{n,n}}{t_{n,n+1}} & -\frac{t_{n,n-1}}{t_{n,n+1}} \\
\mathds{1}&0
\end{array}
\right) 
\left(
\begin{array}{c}
\psi_{n} \\
\psi_{n-1}
\end{array}
\right)
\label{tramat_eqn}
,}
where $h_{n,n}$ is the Hamiltonian of the $n$-th  slice. In a bipartite lattice, the sites can be separated into A (blue circle) and B sites (grey circle) (see Fig.~\ref{rigid_shift}). For the nearest neighbour Hamiltonian defined in Eq.~\ref{ham}, the transfer matrix equation~(\ref{tramat_eqn}) decouples in individual sublattices at $E=0$ and the wavefunction transfer is restricted to one sublattice. The transfer matrix in Eq.~(\ref{tramat_eqn}) gives the wave amplitude at the lattice site $(n+1,b)$ as   
\eq{
\psi_{n+1,b}{=}-\frac{t^{a,b}_{n}}{t^b_{n,n+1}} \psi_{n,a} - \frac{t^{b,c}_{n}}{t^b_{n,n+1}}\psi_{n,c}-\frac{t^b_{n-1,n}}{t^b_{n,n+1}} \psi_{n-1,b}
\label{a1}
}

In particular, for the brickwall (equivalent to honeycomb) lattice the bond $t^b_{n-1,n} = 0$ (see e.g. Fig.~\ref{fig:phasediag}) and the Eq.~\eqref{a1} reduces to 
\begin{eqnarray}
\psi_{n+1,b} = -\frac{t^{a,b}_{n}}{t^b_{n,n+1}} \psi_{n,a} - \frac{t^{b,c}_{n}}{t^b_{n,n+1}}\psi_{n,c}
\label{a2}
\end{eqnarray}
i.e., the transfer only involves two slices $(n,n+1)$ in one iteration.
Here, the bonds are distributed equally between $[0, 1]$ with an extra multiplicative factor $e^{\delta}$ for the horizontal bonds. 
Therefore, the multiplicative factor comes out of the transfer matrix multiplication, which is the rigid shift of the spectrum as seen in Fig.~\ref{fig:lyp_dos_A_sub}. The magnitude of the shift is proportional to $\delta$ and in this sense the exponent is $\nu = 1.0$ by construction as explained in the main text.

In contrast, for the square lattice, the wave amplitude $\psi_{n+1,b}$ involves three slices at one iteration of transfer and the spectrum is non-rigid (not shown here). In this scenario, it is not entirely obvious that $\nu\approx 1$ and therefore, we resort to numerical simulation. 

\section{Finite size scaling analysis}
\label{scaling_appendix}
\begin{table*}
\caption {Finite size analysis for Transfer matrix data of staggered dimerized graphene as shown in Fig~\ref{fig:loclaizedphase_stagger_gphx}(a)}
\begin{tabular}{ p{2cm} p{0.5cm} p{0.5cm} p{0.5cm} p{1cm} p{1cm} p{2.5cm} p{2.5cm} p{1cm} p{0cm}}
 \hline\hline
$L$ & $n_R$ &$n_I$ & $m_R$ & $m_I$&  N & $\delta_c$ & $\nu$ & GOF  \\ 
 \hline

 32-384 & 6 & 0 & 1 & 0 & 10 & $1.430\pm0.002$   & $1.07\pm0.04$ & 0.03 & \\
 32-384 & 6 & 0 & 2 & 0 & 11 & $1.431\pm0.001$   & $1.05\pm0.05$ & 0.04 & \\
 32-384 & 4 & 0 & 1 & 0 & 8 & $1.428\pm0.005$   & $1.13\pm0.05$ & 0.16 & \\
 32-384 & 4 & 0 & 2 & 0 & 9 & $1.431\pm0.002$   & $1.10\pm0.06$ & 0.13 & \\

 \hline

 32-384  & 4 &1& 1 &0& 11 & $1.430\pm0.005$  & $1.04\pm0.04$  & 0.46  &     \\
 64-384  & 4 &1& 1 &0& 11 & $1.430\pm0.003$  & $1.04\pm0.09$  & 0.51  & \\
 128-384 & 4 &1& 1 &0& 11 & $1.431\pm0.003$  & $1.03\pm0.05$  & 0.43  & \\
 
\hline\hline
\label{table1}
\end{tabular}
\caption {Finite size analysis for conductance data in staggered dimerized graphene as shown Fig~\ref{fig:loclaizedphase_stagger_gphx}(c)}
\begin{tabular}{ p{2cm} p{0.5cm} p{0.5cm} p{0.5cm} p{1cm} p{1cm} p{2.5cm} p{2.5cm} p{1cm} p{0cm}}
 \hline\hline
$L$ & $n_R$ &$n_I$ & $m_R$ & $m_I$&  N & $\delta_c$ & $\nu$ & GOF  \\ 
 \hline
 32-112 & 6 & 0 & 1 & 0 & 10 & $1.457\pm0.002$  & $1.11\pm0.04$ & 0.45 & \\
 32-112 & 4 & 0 & 1 & 0 & 8 & $1.456\pm0.002$   & $1.12\pm0.05$ & 0.46 & \\
 \hline
 48-112  & 4 & 1 & 1 & 0 & 11 &$1.455\pm0.003$  & $1.15\pm0.09$  & 0.39  &  \\
 64-112  & 4 & 1 & 1 & 0 & 11 &$1.456\pm0.003$  & $1.13\pm0.07$  & 0.29  &    \\
\hline\hline
\label{table1}
\end{tabular}
\caption{Finite size scaling analysis table for data in Fig~\ref{fig:loclaizedphase_stagger_gphx}. $N$ is the number of parameters to fit. The critical parameters $(\delta_c,\nu)$ are found by fitting the data in Fig.~(\ref{fig:loclaizedphase_stagger_gphx}) to the scaling form in Eq.~(\ref{final_expansion}). The bootstrap error has been obtained by repeated re-sampling (10000 data points) of the original data and the error is the standard deviation of the resulting sampling distribution. The goodness of fit parameter as defined in Eq~\ref{chi_square} is also given in the table.}
\end{table*}
\begin{table*}
\caption {Finite size analysis table for transfer matrix data in plaquette dimerized graphene as shown in Fig~\ref{fig:loclaizedphase_plaquettes}(a)}
\begin{tabular}{ p{2cm} p{0.5cm} p{0.5cm} p{0.5cm} p{1cm} p{1cm} p{2.5cm} p{2.5cm} p{1cm} p{0cm}}
 \hline\hline
$L$ & $n_R$ & $n_I$ & $m_R$ & $m_I$ &  N & $\delta_c$ & $\nu$ & GOF  \\ 
 \hline
 30-384 & 6 & 0 &1  & 0 &10 &$0.005\pm0.002$   & $1.64\pm0.04$ & 0.82 & \\
 30-384 & 4 & 0 &1  & 0 &8  &$0.009\pm0.003$   & $1.58\pm0.04$ & 0.26 & \\
 30-384 & 4 & 0 &2  & 0 &9  &$0.005\pm0.002$   & $1.55\pm0.05$ & 0.76 & \\
 \hline
 66-384  & 4 & 1 & 1& 0 &11 & $0.007\pm0.003$  & $1.54\pm0.06$  & 0.28  &  \\
 126-384 & 4 & 1 & 1& 0 &11 & $0.009\pm0.003$  & $1.59\pm0.09$  & 0.43  &    \\
\hline\hline
\label{table1}
\end{tabular}
\caption {Finite size analysis for conductance data in staggered dimerized graphene as shown Fig~\ref{fig:loclaizedphase_plaquettes}(c)}
\begin{tabular}{ p{2cm} p{0.5cm} p{0.5cm} p{0.5cm} p{1cm} p{1cm} p{2.5cm} p{2.5cm} p{1cm} p{0cm}}
 \hline\hline
$L$ & $n_R$ &$n_I$ & $m_R$ & $m_I$&  N & $\delta_c$ & $\nu$ & GOF  \\ 
 \hline
 32-128 & 6 & 0 & 1 & 0 & 10 & $0.005\pm0.002$   & $1.47\pm0.04$ & 0.24 & \\
 32-128 & 4 & 0 & 1 & 0 & 8 & $0.003\pm0.002$   & $1.45\pm0.05$ & 0.56 & \\
 \hline
 48-128 & 6 & 0 & 1 & 0 & 11 &$0.015\pm0.003$  & $1.45\pm0.09$  & 0.51  &  \\
 64-128 & 6 & 0 & 1 & 0 & 11 &$0.010\pm0.005$  & $1.44\pm0.05$  & 0.15  &    \\
 80-128 & 6 & 0 & 1 & 0 & 11 &$0.009\pm0.003$  & $1.47\pm0.08$  & 0.18  &    \\

\hline\hline
\label{table1}
\end{tabular}

\caption{Finite size scaling analysis for data in Fig.~\ref{fig:loclaizedphase_plaquettes}. The critical parameters $(\delta_c,\nu)$ are found by fitting the data in Fig.~(\ref{fig:loclaizedphase_plaquettes}) to the scaling form in Eq.~(\ref{final_expansion}). The bootstrap error has been obtained by repeated re-sampling (10000 data points) of the original data and the error is the standard deviation of the resulting sampling distribution. The goodness of fit parameter as defined in Eq.~(\ref{chi_square}) is also given in the table.}
\end{table*}

This section summarizes the finite size scaling analysis used to estimate $\nu$ and $\delta_c$ in the localized phase of the model. We follow a approach similar to~\citep{SlevinPRL1999correction,SlevinPRL2001reconcile} and expand the scaling function in the leading relevant ($\chi$) and irrelevant $(\phi)$ scaling variables as follows, 
\eq{
	\frac{\xi}{M} = \mathcal{F}\left( \chi(w)~M^{1/\nu},\phi(w) M^{y} \right),
	w=(\delta_c-\delta)/\delta_c,
}
where $\mathcal{F}$ is a generic scaling function and M is the width of the quasi-1D sample. The irrelevant exponent $y<0$ characterizes correction to finite-size scaling. The scaling variables are expanded in terms $w$ as, 
\eq{
  \chi(w)=\sum_{n=0}^{m_R} a_n w^n,
  \phi(w)=\sum_{n=0}^{m_I} b_n w^n, a_0=0.
  \label{chi_phi_expansion}
}
A further Taylor expansion of $\mathcal{F}$ leads to,
\begin{eqnarray}
	\frac{\xi}{M} = \sum_{m}^{n_R}\sum_{n}^{n_I}a_{mn}\chi^mL^{m/\nu}\phi^nL^{ny}\mathbb{F}_{mn}.
	\label{final_expansion}
\end{eqnarray}
An expansion of Eq.~\eqref{final_expansion} to order $(m_R,m_I,n_R,n_I)$ gives $N= 2 + m_R + m_I + (n_R+1)(n_I+1)$ number of parameters to fit. An analogous expansion also used for the observable  $\lng$ used in the main text.
We use non-linear least squares to fit the function to the available data. In the main text, we kept the order of expansion $(m_R,m_I,n_R,n_I) = (1,0,4,0)$ to avoid over-fitting of the model. But we checked the stability of the critical exponents to various other orders of expansion $(m_R,m_I,n_R,n_I)$ and the results are summarized in table~\ref{table1} along with the goodness of fit (GOF) defined as follows   
\begin{equation}
 \mathrm{GOF} =\sum_{i=1}^{N_D} (F_i - \Gamma_i )^2/ \Gamma_i, 
 \label{chi_square}
\end{equation}
where $F_i$ is the scaling function evaluated at $i^{th}$ data point, $\Gamma_i$ is numerically observed value at the same data point and $N_D$ is the number of such data points. The error bars are estimated using the bootstrap resampling of the data set. The irrelevant exponent $y$ is found to be rather large.

\end{document}